\documentclass{article} 
\usepackage[utf8]{inputenc} 
\usepackage{geometry} 
\usepackage{lmodern}
\usepackage[pdftex]{graphicx} 
\usepackage{wrapfig}
\usepackage{listings}
\usepackage{color}
\usepackage{amsmath}
\usepackage{textcomp}
\usepackage{chngpage}
\usepackage[pdfauthor={Timothy J. Harrison}, pdftitle={Towards in-situ cleaning of a trapped ion quantum computer}, pagebackref=true, pdftex,colorlinks,citecolor=blue]{hyperref}
\usepackage[capitalize]{cleveref}
\usepackage{booktabs} 
\usepackage{array} 
\usepackage{paralist} 
\usepackage{verbatim} 
\usepackage{todonotes}
\usepackage{subfig} 
\usepackage{float}
\usepackage{fancyhdr} 
\usepackage{appendix}
\usepackage{gensymb} 
\usepackage{mathrsfs}
\usepackage{multirow}
\usepackage{caption}
\usepackage{sectsty}
\usepackage[section]{placeins} 
\allsectionsfont{\sffamily\mdseries\upshape} 

\usepackage[nottoc,notlof,notlot]{tocbibind} 
\usepackage[titles,subfigure]{tocloft} 

\newcommand{\nmmin}{nm $\text{min}^{-1}$}
\newcommand{\HRule}{\rule{\linewidth}{0.5mm}}
\def\ifmonospace{\ifdim\fontdimen3\font=0pt }

\usepackage{cite}
\usepackage{units}
\usepackage[parfill]{parskip} 

\begin{document}
\begin{titlepage}

\begin{center}

\includegraphics[width=0.15\textwidth]{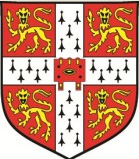}\\[1cm]    

\textsc{\LARGE University of Cambridge}\\[1.5cm]

\textsc{\Large Experimental and Theoretical Physics}\\[0.5cm]
\textsc{\Large Part III Project}\\[0.5cm]

\HRule \\[0.4cm]
{ \huge \bfseries Towards in-situ cleaning of a trapped ion quantum computer}\\[0.4cm]

\HRule \\[1.5cm]

\begin{minipage}{0.4\textwidth}
\begin{flushleft} \large
\emph{Author:} \\
Timothy J. Harrison
\end{flushleft}
\end{minipage}
\begin{minipage}{0.4\textwidth}
\begin{flushright} \large
\emph{Supervisor:} \\
Prof.~Michael K\"ohl
\end{flushright}
\end{minipage}

\vfill
  
\includegraphics[width=0.15\textwidth]{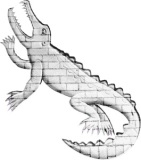}\\[1cm] 
{\large November 2011 - May 2012}

\end{center}

\end{titlepage}

\begin{abstract}
\noindent A plasma glow discharge system was created using a conventional microwave oven to ignite and maintain the plasma. The system was used for plasma cleaning and its properties were analysed to assess its viability for removing surface contaminants, which cause anomalous heating, from Paul trap electrodes used in ion trapping. Qualitative results showed that the argon plasma could remove copper oxide layers formed on thin sheets in 1-2 minutes. The addition of air into the plasma system allowed for the cleaning of more complex hydrocarbon contaminants, as highlighted by the removal of permanent marker pen. The surface removal rate of the system was evaluated by measuring the time taken to remove gold plating from washers and aluminium plates. A nominal rate of \unit[5\text{--}15]{\nmmin} was found under optimal conditions. Plasma treatment, in addition to wet chemical cleaning, was found to increase the finesse of an optical cavity by a factor of two, compared to acetone cleaning alone. These results are promising for the system's in-situ use on Paul trap electrodes and could be a feasible technique for increasing their lifetime. 
\end{abstract}

\section{Introduction}\label{sec:intro}
\subsection{Miniaturisation of ion traps and anomalous heating}\label{subsec:intro:minit}
It was first proposed in 1995 by Cirac and Zoller \cite{R19} that a linear Paul trap, for which Paul won the 1989 Nobel Prize \cite{nobel}, could be utilised to create entangled states for quantum computation. The linear ion trap contains four electrodes driven to produce a radio frequency quadrupole potential with an additional DC field to allow axial trapping .The isolated ions act as a quantum system and can be confined, coherently manipulated and characterised for long periods, whilst experiencing only small environmental perturbations \cite{LVREF1}. The system has been used to successfully implement basic quantum protocols and create entangled states \cite{LVREF2}. Approximately 25 laboratories around the world are investigating quantum information processing using trapped ions. 

Following the trend of classical computation, the community has been driven towards miniaturising such ion traps. Segmented trap electrodes manipulating ion crystals were developed specifically to overcome the difficulties of scalability \cite{LVREF2}. To achieve the high trap frequencies and fast splitting of the ion crystal, miniaturisation of the traps is required and currently ion-surface separations of less than \unit[100]{\textmu m} are being pursued. Unfortunately, the heating rate of trapped ion motion, $\dot{n}$, was found to be anomalously large \cite{R11} compared to that expected from the thermal Johnson noise caused by the electric circuitry.

The most common theoretical model assumes $\dot{n}$ is caused by a fluctuating patch potential on the electrode surface. The resulting noisy electric field at the position of the ion couples to its charge and results in fluctuating forces. One can thus derive \cite{LVREF4} that $\dot{n} \propto d^{-4}$ and the spectral density of the noise, $S(\omega) \propto \omega^{-1}$, where $d$ is the ion-surface separation and $\omega $ the frequency. Recent models, using more complex geometries have built upon these results \cite{R20}. The $d$ dependence of $\dot{n}$ was first measured by Deslauriers et al. in 2006 \cite{LVREF3}. They measured $\dot{n} \propto d^{-3.5\pm 0.1}$ and showed cryogenic cooling reduced $\dot{n}$ by an order of magnitude. These results suggest that the microscopic noisy potentials are thermally driven and they are thought to be caused by surface contaminants on the electrodes; this is supported further by the observed increase in heating rates over time \cite{TimeIncrease}. The observed noise could be caused by merely one monolayer of impurity adsorbate atoms on the ion trap surface.

\subsection{Plasma cleaning and microwave ignition}\label{subsec:intro:plasma}
The term \textit{plasma} was used first by Langmuir in 1929 to describe the ionised gases produced in the development of vacuum tubes under large currents \cite{LangmuirFist}. It is now more rigorously defined as a quasi neutral collection of charged and neutral particles characterised by a collective behaviour. The nature of the plasma and its possible applications depend dramatically and abstrusely on parameters such as temperature and pressure. The glow discharge process of a plasma is characterised by electron densities $\sim\unit[10^{14}\textrm{--}10^{18}]{m^{-3}}$ and electron energies $\sim\unit[1\textrm{--}15]{eV}$ \cite{plasmaBook1}. This regime was first used to clean the inside of vacuum chambers during the 1970s, notably at CERN in 1977 \cite{LVREF5}. Plasma cleaning has since been used in numerous industries including semiconductor processing\cite{ApplicationSemiconductor}, metallurgy\cite{LVREF18} and medical technologies\cite{ApplicationDental}. 

The electrons, ions and radicals, generated in a plasma, clean solid surfaces via three interactions: heating, etching and sputtering. Heating can slowly remove physiosorbed or lightly bound contaminants, most notably water. Etching occurs when atoms or radicals from the plasma chemically react with the surface. Adsorbed species can react with contaminants and the more volatile products may desorb. Etching rates can be very fast and chemically selective \cite{plasmaEtching}. Sputtering occurs when collisions of active plasma species with contaminants lead to their removal from the surface. Sputtering is non-selective, but depends greatly on the nature of the surface and contaminant. It also can cause ion implantation in the surface and structural changes\cite{LVREF7}. Furthermore, redeposition of sputtered material may simply redistribute contaminants about the surface, rather than remove them.

The most common sources \cite{LVREF7, LVREF15, LVREF16, LVREF17} for plasma cleaning are DC and Radio Frequency (RF) glow discharges at low pressures. In a DC system, a substrate is used as an electrode of the system and the strong ion bombardment causes surface defects to occur \cite{LVREF15}. RF cleaners are widely used for cleaning a range of metals and dielectrics and are the most common commercial device \cite{LVREF7}. Microwave (MW) sources are less commonly found, but are more efficient in terms of having a higher density of radicals per Watt, leading to a higher removal rate \cite{MWPlasmaAdvantages}. 

The idea of using a conventional MW oven as the source for plasma ignition was first patented by Ribner \cite{USPatent,MWBookMehrdad} in 1989 and a few such models are still sold commercially \cite{MWPlasmaAdvantages}. A MW oven uses a magnetron to generate electromagnetic radiation of frequency of \unit[2.45]{GHz} which feeds into a metallic chamber via wave guides. Any substance absorbs a fraction of these MWs depending on its dielectric constant. If the device's power (typically \unit[600--800]{W}) is sufficient, a plasma ignition of the gas inside may occur. MW plasma cleaners have been shown to have some advantages over more conventional RF models \cite{MWPlasmaAdvantages}. They have faster cleaning rates and do not cause damage to microelectronic devices. A system using a MW oven is, of course, cheaper than a commercial RF or MW model, which usually cost more than \pounds 10,000.

These reactive plasma cleaning techniques have yet to be adapted to the complex geometries and materials of ion traps. Preliminary experiments \cite{LVPROJ} showed that a DC plasma was unsuitable due to redeposition of sputtered material. This project aims to characterise the properties of a plasma cleaning system built from a conventional MW oven. Furthermore, it aims to assess whether this system is feasible for removing surface contaminants from Paul trap electrodes, thus reducing the observed anomalous heating. This system could eventually be built around the ion trap to allow in-situ cleaning. In \cref{sec:theoretical} an overview the theory of MW absorption and the criteria for plasma ignition is derived. The method, results and discussion are put forward in \cref{sec:method,sec:results,sec:discussion} respectively. \cref{sec:conclusion} contains the overall conclusions. 

\section{Theoretical Background}\label{sec:theoretical}

Since their first observation by Sir William Watson in 1748, electrical breakdown of gases have been researched. During the later half of the nineteenth century and much of the twentieth century, understanding of plasma glow discharges grew rapidly. Whilst DC breakdowns were relatively well understood by this period, analysis of high frequency breakdowns were not investigated even into the 1930s. The development of radar during World War II led to the rapid improvement of MW technology\footnote{The MW oven was invented Leonard Gfoellner in 1947}, as highlighted by the flurry of papers on the topic of MW breakdown of gases in the late 1940s \cite{mw49,mwHerlin1,mwHerlin2,mwHerlin3,mwHerlin5,mwHerlin6}. The analysis of such high frequency breakdowns are considerably more involved than their DC counterparts due to the many oscillations of an electron between kinetic collisions. The aim of this section is to show that a conventional MW oven can supply enough power to cause a plasma glow discharge to exist in an argon vessel as used in this experiment. For a complete discussion of \textit{Microwave Breakdown in Gases} see the work of MacDonald \cite{MwBreakdownBook}, and for more recent developments and techniques \cite{R21} is suggested.

\subsection{The criterion for plasma glow discharge at microwave frequencies}\label{subsec:breakdownCriterion}

We begin by considering a lone free electron in a vessel of gas exposed to electromagnetic radiation of frequency $\omega$. The equation of motion due to the Lorentz force is given by:

\begin{equation}\label{eq:ofmotion}
\frac{d^2\mathbf{r}}{d t^2}=-\frac{e}{m_\text{e}} (\mathbf{E} + \mathbf{v} \times  \mathbf{B})
\end{equation}

where $m_\text{e}$ is the electron mass, $\mathbf{r}$ is the vector position of the electron relative to an arbitrary point, $\mathbf{v}$ is the velocity of the electron, $t$ is time and $\mathbf{E}$ and $\mathbf{B}$ are the (time dependent) electric and magnetic fields respectively. The second term in \cref{eq:ofmotion} is neglected as  $\left|\frac{\mathbf{E}}{\mathbf{v} \times \mathbf{B}}\right| \gg 1$ for the electron speeds considered. From \cref{eq:ofmotion}, it is easily shown that under such radiation the electron oscillates with its velocity $\sim\unit[\nicefrac{\pi}{2}]{radians}$ out of phase with the fields, and so no power is taken on average from the applied field. Hence, the electron only attains energy from the field by undergoing collisions with gas atoms present in the vessel. In this process its ordered oscillatory motion is changed to random motion upon collisions, which can be considered elastic as $m_\text{e}\ll m_{\text{atom}}$. At sufficiently small $\omega$, the collision frequency, $\nu_\text{c}$, is low enough that the drift current of electrons oscillates with the field and the energy transfer is efficient between collision (this tends to a DC system as $\omega \rightarrow 0$). However, if $\omega \gg \nu_c$, then there are many oscillations of the field per collision and the inertia of electrons causes the out of phase velocity discussed above. Thus, the energy transfer process becomes much less efficient. The energy transfer efficiency can be accounted for by using an effective E-field strength which can be shown\cite{mwEffectiveField,mwHerlin2} to have the form $E_{\text{eff}}^2 = E^2 \frac{\nu_{\text{c}}^2}{\nu_{\text{c}}^2+\omega^2}$. 

The formal condition for a gas discharge breakdown is that the average rate of gain of electrons from ionisation of atoms is equal to (or greater than) the rate of loss of electrons. For the system considered, the principal method of electron withdrawal is loss to the vessel walls via diffusion\footnote{Other electron loss processes include electrostatically coupled (ambipolar) diffusion, attachment of electrons to neutral atoms or molecules and recombination with ions. Furthermore, there could be secondary emission of electrons from surfaces. These processes are harder to treat rigorously and may be more relevant in a gas mixture plasma. Consult  \cite{MwBreakdownBook} for a detailed discussion of secondary loss and gain effects.}. An equivalent expression of the breakdown criterion is that the number of collisions required, $N_{\text{i}}$, before the electron gains a kinetic energy equal to the ionisation energy of an atom, $\unit[U_\text{i}]{eV}$, is smaller than the number of collisions, $N_\text{d}$, before the electron will be lost by diffusion to the walls of the vessel. By integration of \cref{eq:ofmotion} we find that:

\begin{equation}\label{eq:deltaV}
\left\langle \Delta v^2\right\rangle =\left\langle \left(\frac{d r}{d t}\right)^2\right\rangle =\left(\frac{e E_{\text{eff}}}{m_\text{e}}\right)^2\left\langle \tau^2\right\rangle
\end{equation}
where $\left\langle \Delta v^2\right\rangle$ is the average increment in velocity squared per collision due to gain from the field and $\tau = \nicefrac{1}{\nu_\text{c}}$ is the mean free time. The inefficiency of energy transfer has been incorporated by the use of $E_{\text{eff}}$ rather than $E$. Noting that $\left\langle \tau ^2\right\rangle ^{\frac{1}{2}}  = \nicefrac{2}{\nu_\text{c}}$, \cref{eq:deltaV} can be used to attain the average change in energy per collision, $\Delta u$:

\begin{equation}\label{eq:DeltaU}
\Delta u =\unit[\left\langle \frac{1}{2}m_\text{e} \Delta v^2\right\rangle =\frac{1}{m_\text{e}}\left(\frac{e E_{\text{eff}}}{\nu _\text{c}}\right)^2=\frac{e E_{\text{eff}}^2}{m_{\text{e}}{\nu_\text{c}}^2}]{eV}
\end{equation}

\begin{equation}\label{eq:Nionisation}
\implies N_\text{i}=\frac{m_\text{e} U_\text{i} \nu_c^2}{e E_\text{eff}^2}
\end{equation}

Modelling the electron's movement as a random walk with a Maxwell-Boltzmann distribution of velocities and mean free path, $l$, leads to a relationship between the mean square distance diffused $\left\langle R^2\right\rangle$ and the number of collisions $N$:

\begin{equation}\label{eq:randomWalk}
\left\langle R^2\right\rangle =\frac{2 l^2}{3}N \implies N=\frac{3\left\langle R^2\right\rangle}{2 l^2}
\end{equation}

To relate $N$ to $N_\text{d}$ we must calculate the $\left\langle R^2\right\rangle$ which represents the average distance for a particle to reach the boundary of the vessel. This is known as the diffusion length, $\Lambda$, and replacing $\left\langle R^2\right\rangle$ with $\Lambda^2$ gives $N_\text{d}$:

\begin{equation}\label{eq:Ndiff}
N_\text{d}=\frac{2\Lambda^2}{3l^2}
\end{equation}

Therefore, our condition for breakdown can be expressed:

\begin{equation}\label{eq:breakdownCriterion}
\frac{N_\text{i}}{N_\text{d}}=\frac{3 l^2 m_\text{e} U_\text{i} {\nu_c}^2}{2 e \Lambda ^2 {E_{\text{eff}}}^2} = \frac{3 l^2 m_\text{e} U_\text{i} \left(\omega^2+{\nu_\text{c}}^2\right)}{2 e E^2 \Lambda ^2} \leq 1
\end{equation}

\subsection{Numerical evaluation of breakdown criterion for our experimental arrangement} \label{subsec:numericalTheory}

Now it remains to ensure that the above criterion will be satisfied in our experiment, noting that any satisfaction of \cref{eq:breakdownCriterion} will lead to a runaway breakdown and rapidly produce a plasma glow discharge. Firstly the values\footnote{value obtained from National Institute for Standards and Technology} of $U_\text{i} = \unit[15.76]{eV}$ and $\omega = 2\pi\times\unit[2.45]{GHz}$ are known for our vessel of argon gas in a MW oven. To calculate $\Lambda$ rigorously, one must find a solution for the diffusion equation of electrons in our approximately hemispherical vessel in the presence of a source and a sink (namely our gas inlet and vacuum); this is complicated further by the presence of the background gas and the treatment requires analysis of several diffusion modes\cite{MwBreakdownBook}. Here we approximate the system as cylindrical and neglect the effect of the source and sink to obtain an order of magnitude approximation, $\Lambda \sim \left(\nicefrac{d}{\pi}\right)$ where $d\sim\unit[23]{cm}$ is the diameter of the vessel. It can thus be seen from \cref{eq:breakdownCriterion} that it is easier to ignite a plasma in a larger vessel as it will take longer for an electron to diffuse to the walls. The mean free path of an electron is obtained from kinetic theory:

\begin{equation}\label{eq:mfpCorrected}
l=\frac{k_\text{b} T}{\sqrt{2}P \sigma}
\end{equation}

where $P$, $T$ and $\sigma$ represent the pressure, temperature and $\sigma \approx \unit[1.1\times 10^{-19}]{m^2}$ is the empirically found cross section for collision between an electron and argon atom \cite{ElectricalDischargesBook}. Kinetic theory is used to estimate the electron velocity and from \cref{eq:mfpCorrected} one can easily estimate the collision frequency,

\begin{equation}
\nu_c=\frac{4P \sigma}{\sqrt{m_\text{e}\pi k_\text{b} T}}
\end{equation}

Finally we assert that $E$ in an \unit[800]{W} conventional MW oven varies from \unit[800-5000]{$\text{V m}^{-1}$} and can be substantially smaller in areas of deconstructive interference \cite{MWBookMehrdad, MWBookApplications}. Utilising the above values and assuming $T = \unit[300]{K}$, $P = \unit[1]{mbar}$ and $E=\unit[3000]{V m^{-1}}$ we find that $\nicefrac{N_\text{i}}{N_\text{d}} \approx 0.02$ and hence we expect a glow discharge breakdown. More generally we find:

\begin{equation}
\frac{N_\text{i}}{N_\text{d}}=\frac{2.6 \times 10^{-17} T}{E^2} + \frac{20800 T^2}{E^2 P^2} \approx \frac{20800 T^2}{E^2 P^2}
\end{equation} 

\begin{figure}[b!]
\centering
\includegraphics[width=\textwidth]{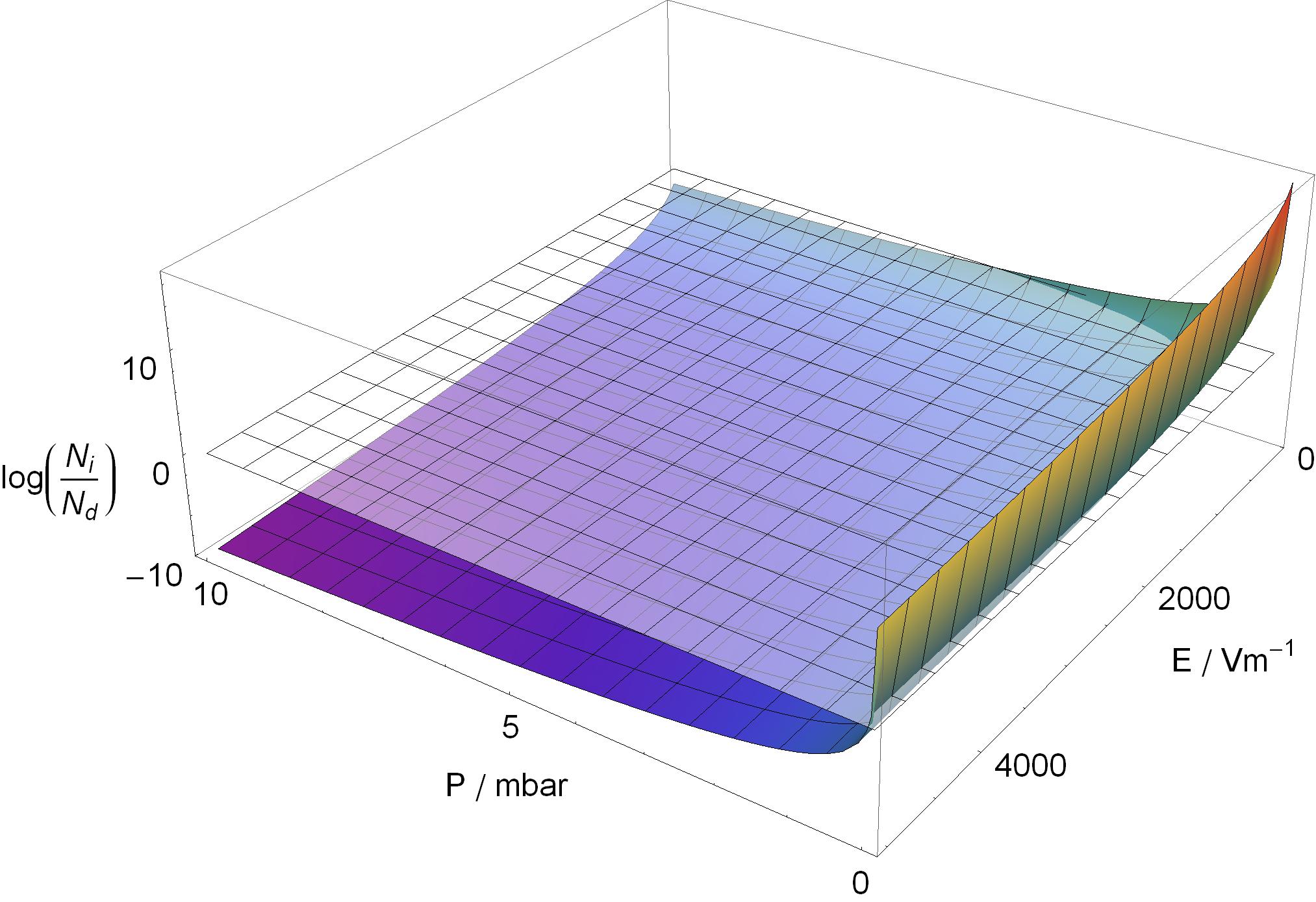}
\caption{Theoretical condition for breakdown. For plasma ignition we predict that $\nicefrac{N_\text{i}}{N_\text{d}}\le 1\implies \log\left(\nicefrac{N_\text{i}}{N_\text{d}}\right) \le 0$. The graph shows the theoretical value of this expression for varying $E$ and $P$ that may exist in the MW oven and assumes approximate values for the other parameters. The flat translucent plane represents the critical value above which we expect no plasma breakdown to occur. Note how plasma is not ignited at sufficiently low $P$ and $E$. When $P$ is too low an electron will collide with the vessel walls too often and when $E$ is too low it will not gain enough energy per atomic collision. This model makes several assumptions, preeminently, the assumption that $\omega \gg \nu_c$ breaks down at high $P$, where rapid liberation of electrons (Townsend avalanche) brings us into the arcing regime \cite{glowToArc,glowtoArc2,glowToArc3}.}
\label{fig:critcalValSurf}
\end{figure}

It is unlikely that T will vary by more than a factor of 2, whereas, $E$ and $P$ may vary by orders of magnitude during the experiment. \cref{fig:critcalValSurf} shows how the value of $\nicefrac{N_\text{i}}{N_\text{d}}$ varies with $E$ and $P$, and hence in what regions we expect plasma glow breakdown to occur.

\subsection{Plasma parameters}\label{subsec:plasmaPararameters}

To conclude \cref{sec:theoretical}, some useful properties of plasmas are defined. For a more detailed discussion and derivation, consult any comprehensive book on plasma physics, for example \cite{plasmaBook1}. 

A travelling wave propagating through a plasma with wave vector $\mathbf{k}$ can be represented as $E e^{-\mathrm{i} \left( \omega t - n \left( \omega \right) \mathbf{k} \cdot \mathbf{r} \right)}$. Using \cref{eq:ofmotion} it is easy to show that the frequency dependent refractive index is:

\begin{equation}\label{eq:plasmaFreq}
n(\omega )=\sqrt{1-\left(\frac{\omega_p}{\omega }\right)^2}
\end{equation}

where the plasma frequency, $\omega_\text{p}=\sqrt{\frac{n_\text{e} e^2}{m_\text{e}\epsilon_\text{0}}}$. Thus, if $\omega <\omega_\text{p}$ the wave will be evanescent\footnote{This can be related to the Debye length, which is the scale over which electrons screen out electric fields in a plasma.} with a decay constant,
\nopagebreak
\begin{equation}\label{eq:decayConstant}
\alpha =k\sqrt{\left(\frac{\omega_\text{p}}{\omega }\right)^2-1}
\end{equation} 
\pagebreak
\section{Method}\label{sec:method}
\subsection{Microwave oven customisation and plasma generation}
The initial task of this experiment was to customise the MW oven and make it suitable for the stable generation of a plasma glow discharge. Details of the components used and unexpurgated instructions on the adaptations are given in \ref{app:MWCustomisation} and can be used to recreate our system exactly. 

\begin{figure}[ht]

\centering
\subfloat[][Baseplate]{\label{fig:basePlate}\includegraphics[width=0.4\textwidth]{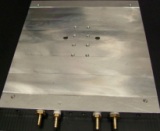}}
\hspace{0.01\textwidth}
\subfloat[][Side piping]{\label{fig:MWAdaptations}\includegraphics[width=0.4\textwidth]{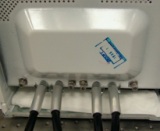}}
\hspace{0.01\textwidth}
\subfloat[][Internal chamber]{\label{fig:CompleteSystem}\includegraphics[width=0.4\textwidth]{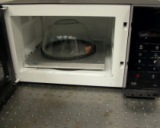}}
\caption{Adaptations made to the conventional MW oven to allow it to house a plasma glow discharge. Refer to \ref{app:MWCustomisation} for a complete description of the manufacturing process. The baseplate was water-cooled to \unit[7]{\degree C} (the outermost hose connectors) and two additional paths allowed for gas flow in and out (the large central holes and central hose connectors). \cref{fig:MWAdaptations} shows how the hose connectors passed through the side of the MW oven and the visible screws secure and ground the baseplate. \cref{fig:CompleteSystem} shows the entire system with the glass container.}
\label{fig:construction}
\end{figure}

A water-cooled, aluminium baseplate was manufactured to lay inside the cooking chamber of the \unit[800]{W} conventional MW oven. Paths through the baseplate allowed for gas to flow in and be removed (\cref{fig:basePlate} ). Holes drilled in the left hand side of the MW (\cref{fig:MWAdaptations} ) were used to connect the baseplate to external gas sources, a vacuum pump and a water cooling system. The gas inlet was connected with a T-junction to two flowmeters which allowed the admission of argon and air into the system at individually controlled rates. Additional holes were used to secure and ground the baseplate to the MW oven. Inverted glass containers of various sizes were placed on the baseplate to act as the vacuum chamber in which the plasma was formed (see \cref{fig:CompleteSystem} ) and pressures down to \unit[$\sim 5\times 10^{-3}$]{mbar} were attainable.

The plasma glow discharge was formed by first pumping the system down and then using the flowmeters to allow argon and/or air into the evacuated region, which caused the pressure to increase (see \cref{fig:pressureFlow} ) to \unit[1--10]{mbar}. The flowmeters measured flow rates between 1 and 10 standard cubic feet per hour (SCFH). A MW oven cooking cycle can be initiated for a chosen amount of time; the plasma should be formed instantaneously and will be visible by a bright purple glow. 

\subsection{Plasma treatments of copper oxide and marker pen}

After the system had been assembled, it was possible to examine its effects on substrates which could have a surface contaminant layer removed. For preliminary and qualitative results we wanted this cleaning effect to be visible. The simplest test was done by drawing a grid-like structure on the inside of the glass chamber with either a dry wipe or permanent marker pen and seeing if the plasma would remove these marks from the glass surface. The efficacy of this removal was examined for varying flow rates and gas combinations. 

Copper oxide layers (\unit[$\sim 1\text{--}10$]{nm} \cite{LVREF8}) form naturally on a sheet of copper left in ambient air and thicker colourful layers could be created by applying a heat gun to a copper sheet for several minutes. These contaminated copper sheets could then be position on the baseplate in a variety of positions and orientations and subjected to the plasma at various flow rates and gas compositions. 

\subsection{Measuring the removal rate of the plasma system}\label{subsec:methodEtch}

The removal rate of the plasma cleaning system was determined in two ways. Firstly, eight \unit[3]{cm} by \unit[3]{cm} aluminium plates were coated with a gold layer of thickness \unit[$\simeq 1$]{\textmu m} by \textit{Modern Metallic Finishes Ltd}. The thickness could be measured accurately using the technique of X-ray fluorescence (XRF) \cite{XRF}. These plates were exposed to plasma in fixed locations under different conditions of flow rates and gas combinations for different lengths of time. Thus by sending the plates back to \textit{Modern Metallic Finishes Ltd} who measured the thickness of the gold layer after treatment, it was possible to calculate how the removal rate varied for different conditions. For these treatments the plate was held between a small glass dish and the inside of the glass chamber. This allowed for all the plates to be kept vertical and adjacent to the glass region where the plasma was brightest. Using this technique the location of different plates was constant to the within \unit[1\text{--}2]{cm}. In all runs the initial pressure of \unit[$(1\pm 0.5) \times 10^{-2}$]{mbar}. Under each set of conditions a separate plate was treated for 18, 36 and \unit[54]{min}.

\begin{figure}[ht]
\centering
\includegraphics[width=0.5\textwidth]{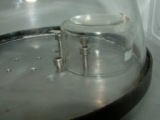}
\caption{A gold-plated washer positioned to be in the brightest region of the plasma. Note that by using a glass dish to raise the washer, the effects of water cooling are lost. Since the washers were much smaller than the aluminium squares, it was possible to expose them to much higher flow rates of argon and air at which the plasma became less uniform.}
\label{fig:goldEtchExperiment}
\end{figure}

In addition to the plates, a similar experiment was performed using gold-plated washers that are provided with SMA connectors. These are quoted as having a gold layer thickness of \unit[76]{nm} \cite{goldPlatedWasher}. This is too small to measure accurately using XRF, but early experiments revealed that this gold layer could be removed in a reasonable timescale (\unit[$\approx 10$]{min}) to expose the brass body, which rapidly formed a colourful oxide in the plasma. Hence, by plasma treating the washers and examining their appearance at regular intervals (\unit[30]{sec}) it was possible to estimate a time at which approximately \unit[76]{nm} had been removed. This was done under various different flow rates and gas combinations. The washers were rested atop a glass dish (\cref{fig:goldEtchExperiment} ) to position them at the height of the brightest region. Due to their smaller size, plasma could be considered uniform over the washers at much higher flow rates than was possible with the larger plates. A washer was defined as cleaned when more than half its surface had the gold completely removed. The results were collated to show variation in the removal rate with the flow rate of argon and presence of air.

\subsection{Plasma treatment of an optical cavity}\label{subsec:methodFinesse}

\begin{figure}[ht]
\centering
\includegraphics[width=0.8\textwidth]{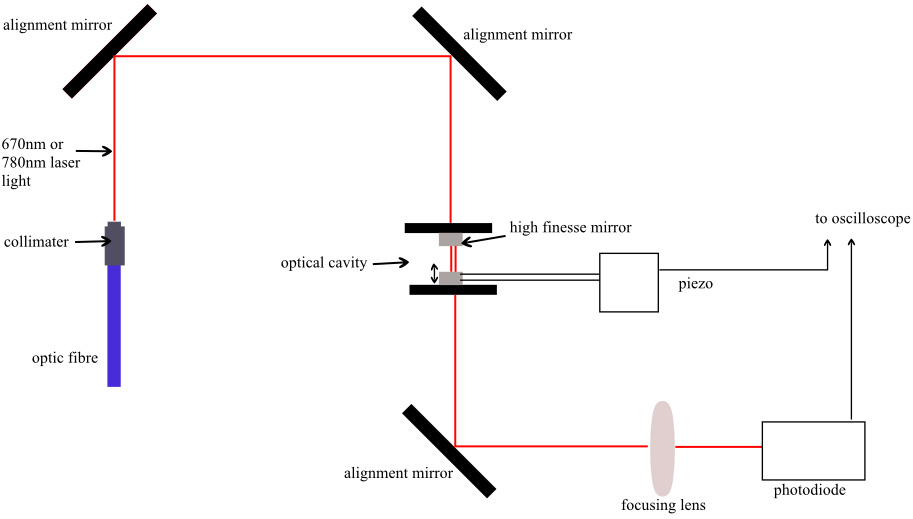}
\caption{Schematic diagram of the setup used to measure the finesse of the optical cavity. An optic fibre and collimator is used to channel laser light of \unit[$\lambda=780$]{nm} (or \unit[$\lambda=670$]{nm} used for initial alignment). Two alignment mirrors are used to ensure the light entering the cavity has the correct orientation and the high finesse mirrors must be accurately positioned themselves. A third alignment mirror then directs the transmitted light through a lens which focuses on a photodiode. The piezo is used to oscillate one mirror rapidly.}
\label{fig:mirrorSetup}
\end{figure}

The final quantitative experiment aimed to assess if plasma cleaning was beneficial for optical devices. The finesse of an optical cavity, $\mathscr{F}$, is a measure of the average number of round trips made by a photon before it is lost or absorbed in the cavity. By considering the absorption coefficient as related to the linewidth, it can be shown\footnote{Absorption coefficient $\kappa = \Delta \nu = \frac{1}{\mathscr{F} \nicefrac{2 d}{c}} \implies \mathscr{F} = \frac{\nicefrac{c}{2d}}{\Delta \nu} = \frac{\delta \nu}{\Delta \nu}$, where $d$ is the mirror separation.} that $\mathscr{F} = \nicefrac{\delta \nu}{\Delta \nu}$ where $\delta \nu$ and $\Delta \nu$ are the peak separation and peak width, respectively. Thus $\mathscr{F}$ quantifies the quality of the optical cavity and is degraded by contamination and surface defects. The system was set up as shown in \cref{fig:mirrorSetup} and was aligned to maximally excite the lowest order Laguerre-Gaussian mode \cite{LagGauss}. At a wavelength of \unit[$\lambda=780$]{nm} the manufacturers documentation states that $\mathscr{F} \approx 30,000$ and for the wavelength used during alignment (\unit[$\lambda=670$]{nm}), $\mathscr{F} \approx 30$. With \unit[$\lambda=780$]{nm} the resonance was incredibly sensitive to perturbations and several measurements were made using the digital oscilloscope and averaged to give a mean value and error. The finesse was measured when one of the mirrors had been dirtied, cleaned with dry acetone, and finally plasma cleaned. This was done once for \unit[$\lambda=670$]{nm} and \unit[$\lambda=780$]{nm}, and at \unit[$\lambda=670$]{nm} an additional plasma cleaning was done to see if further plasma exposure had an effect. The mirror was dirtied by applying a layer of vacuum lubricant oil (\unit[$\sim 1$]{mm}) onto the mirror using a cotton bud. The mirror was cleaned with acetone by removing it from its stand and applying the dry acetone and using lens cleaning tissue. The plasma cleaning routine involved exposure to the plasma with an air flow to reach \unit[2]{mbar} and an additional argon flow of \unit[2]{SCFH} for \unit[250]{sec}. The mirror was positioned on the right hand side of the chamber in a bright spot of the plasma. It should be noted that the untreated state before measurement at \unit[$\lambda=780$]{nm} was no different to the state after plasma cleaning for the experiment at \unit[$\lambda=670$]{nm} which had been conducted before. The mirror attached to the piezo (see \cref{fig:mirrorSetup} ) was not treated throughout the experiment.

\section{Results}\label{sec:results}
\subsection{Observations of the plasma system}

At any pressure above \unit[$5 \times 10^{-3}$]{mbar} and below a cut off of approximately \unit[10]{mbar} a plasma was generated immediately upon exposure to MW radiation. The system allowed for air, argon or a combination of both to be the ionised gas forming the plasma, with multifarious results in each case.

\begin{figure}[ht]
\centering
\subfloat[][air-argon]{\label{fig:argonAirPlasma}\includegraphics[width=0.4\textwidth]{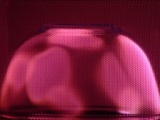}}
\hspace{0.05\textwidth}
\subfloat[][pure argon]{\label{fig:pureArgonPlasma}\includegraphics[width=0.4\textwidth]{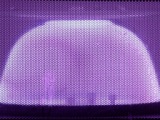}}
\caption{Appearance of an air-argon and pure argon plasma formed in the MW oven. The air-argon plasma has a pressure of approximately \unit[4]{mbar} and the pure argon of \unit[1]{mbar}. Note the dramatic difference in colour and the appearance of irregular patches when air is additionally introduced. In both cases the plasma is brightest nearest the MW source (right hand side) and forms a thin sheath on the inner wall of the glass container.}
\label{fig:plasmaColours}
\end{figure}

A plasma containing exclusively air was quite dim and only formed in a small region of the chamber with a purple colour. At higher flow rates of air it moved away from the glass edge and had small elements of green and orange colour. At very low pressures the air plasma took on a grey appearance. When argon was added in conjunction with air, the plasma became significantly brighter whilst maintaining its purple colour. At all pressures above \unit[1]{mbar} the argon-air plasma formed patches of irregular shapes in a thin layer on the inner surface of the glass chamber (see \cref{fig:argonAirPlasma} ). The patches become less numerous at higher flow rates and were brightest nearest the magnetron source. A pure argon plasma had a distinctly blueish colour (see \cref{fig:pureArgonPlasma} ) and also formed a uniform thin layer around the inner surface of the glass chamber. 

\begin{figure}[ht]
\centering 
\includegraphics[width=0.75\textwidth]{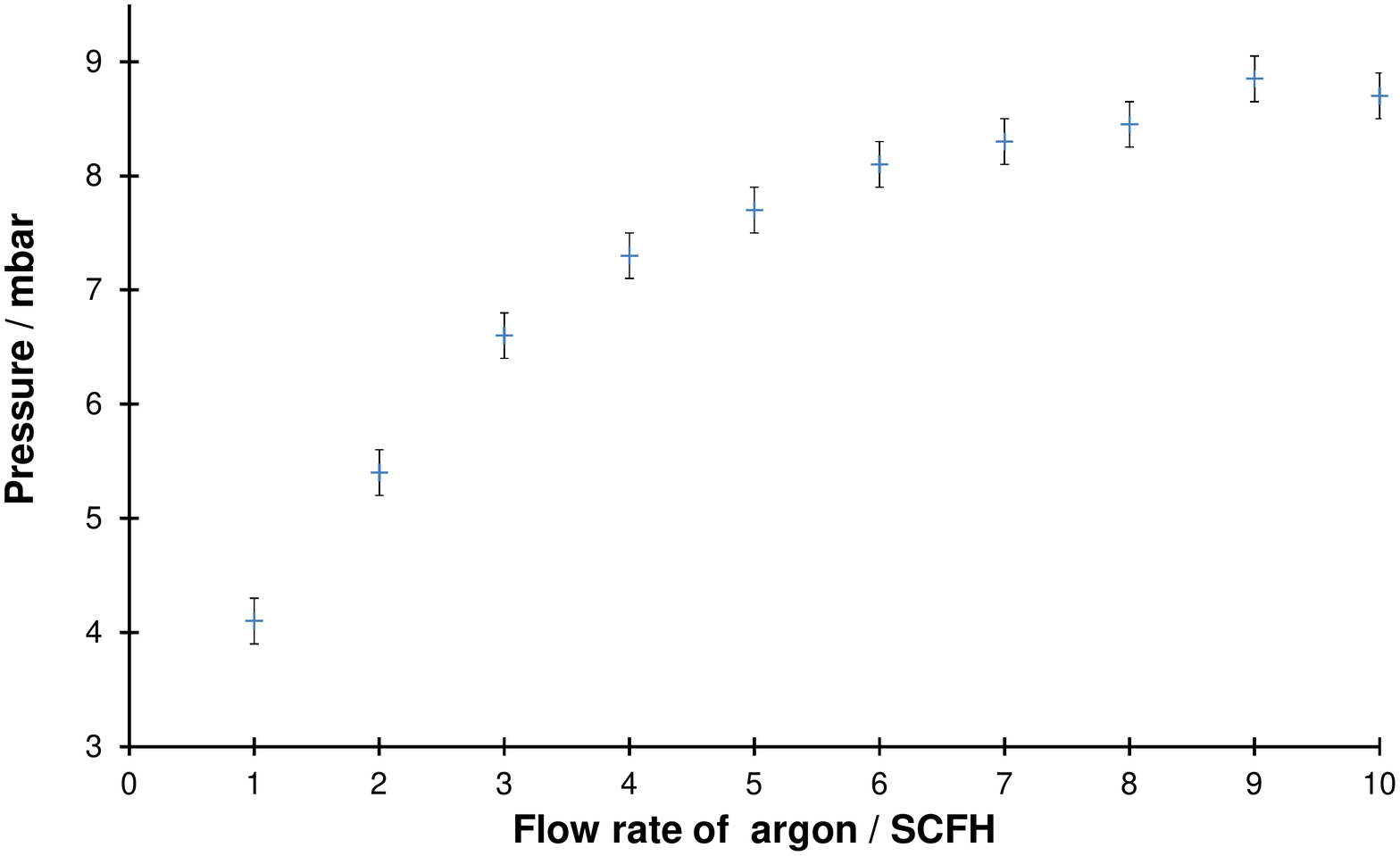}
\caption{Variation of pressure within the chamber with flow rate of argon. The pressure was measured using a Pirani gauge. The accuracy of flow rate measurements was \unit[$\pm 0.2$]{SCFH} due to the limited resolution of the flowmeter scale. As the flow rate increases the pressure increases until a saturation point at around \unit[6]{SCFH}. It was also observed that upon plasma ignition the pressure rapidly increased by as much as a factor of 2 before falling upon stable generation.} 
\label{fig:pressureFlow}
\end{figure}

\begin{figure}[ht]
\centering
\subfloat[][0.05 mbar]{\label{subfig:0p05}\includegraphics[width=0.3\textwidth]{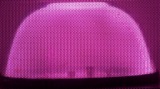}}
\hspace{0.05\textwidth}
\subfloat[][0.5 mbar]{\label{subfig:0p5}\includegraphics[width=0.3\textwidth]{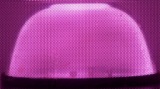}}
\hspace{0.05\textwidth}
\subfloat[][1.5mbar]{\label{subfig:1p5}\includegraphics[width=0.3\textwidth]{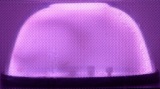}}
\hspace{0.05\textwidth}
\subfloat[][2.5 mbar]{\label{subfig:2p5}\includegraphics[width=0.3\textwidth]{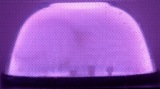}}
\hspace{0.05\textwidth}
\subfloat[][3.0 mbar]{\label{subfig:3p0}\includegraphics[width=0.3\textwidth]{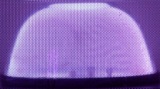}}
\hspace{0.05\textwidth}
\subfloat[][7.0 mbar]{\label{subfig:7p0}\includegraphics[width=0.3\textwidth]{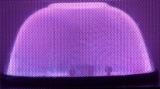}}
\caption{The change in appearance of a pure argon plasma at increasing flow rates (or pressures) of argon. This highlights that the appearance of the plasma changes as the flow rate changes. The initial base pressure of air was \unit[$6\pm 1 \times 10^{-3}$]{mbar}. }
\label{fig:pressuresPlasma}
\end{figure}

The flow rate of gas is correlated to the pressure in the system as shown by \cref{fig:pressureFlow} and the appearance of the plasma changes for different flow rates (or pressure) as demonstrated in \cref{fig:pressuresPlasma}.

\begin{figure}[t]
\centering
\includegraphics[width=0.5\textwidth]{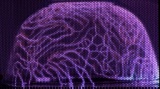}
\caption{Labyrinthine pattern of an air-argon plasma that is formed at high flow rates of argon. The lines cling to the inner glass wall for all but the highest flows rates (\unit[$\geq 8$]{SCFH}) at which point the lines emanate from the wall to central regions of the baseplate. In a pure argon plasma the lines occur at a much lower threshold, are smoother, move less erratically and form a more regular pattern. Bright points are seen where the lines meet the baseplate.}
\label{fig:electricLines}
\end{figure}

For both argon-air and argon plasma there is a dramatic change in appearance at a critical flow rate. At this point the plasma changes to a brighter, bluer colour and forms a labyrinthine pattern that moves rapidly and erratically around the inner glass (\cref{fig:electricLines} ). This occurs at \unit[$\leq 2$]{SCFH} for a pure argon plasma and \unit[4-6]{SCFH} for pure argon when air is allowed to flow in to create a starting pressure of \unit[1-2]{mbar}.

\FloatBarrier
\subsection{Qualitative observations of cleaning on copper oxide and marker pen}

\begin{figure}[ht]
\centering
\subfloat[][natural copper]{\label{subfig:copperNatural}\includegraphics[width=0.4\textwidth]{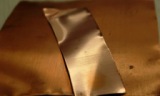}}
\hspace{0.05\textwidth}
\subfloat[][purposefully oxidised]{\label{subfig:copperPurpose}\includegraphics[width=0.4\textwidth]{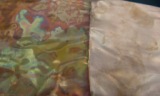}}
\caption{Photographic results of copper oxide removal. Copper sheets with either naturally or purposefully oxidised layers were exposed to a pure argon plasma with a flow rate of \unit[1--2]{SCFH}. After \unit[1--2]{min} of exposure the copper oxide layer had been removed to reveal a paler and shinier copper surface. On average the purposefully oxidised layers (noticeable by the colourful surface) took slightly longer to remove. The cleaning rate was not uniform with some regions being fully cleaned and others still oxidised after short exposures.}
\label{fig:copperCleaning}
\end{figure}

\begin{figure}[ht]

\centering
\subfloat[][uncleaned]{\label{subfig:uncleanedPen}\includegraphics[width=0.4\textwidth]{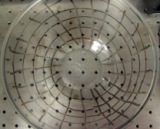}}
\hspace{0.05\textwidth}
\subfloat[][plasma exposure]{\label{subfig:plasmaExposurePen}\includegraphics[width=0.4\textwidth]{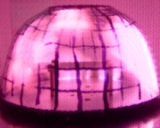}}
\hspace{0.05\textwidth}
\subfloat[][cleaned]{\label{subfig:cleanedPen}\includegraphics[width=0.4\textwidth]{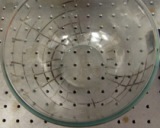}}
\caption{\footnotesize{Removal of marker pen from chamber walls using an air-argon plasma. A grid like structure was drawn on the inside of the chamber using a dry wipe marker pen. Plasma exposure for \unit[90]{s} gave the result shown with marker pen having been removed in the bright regions of the plasma. Further exposure increased the area that was cleaned but some regions where the plasma did not appear active remain marked even after significantly longer exposures.}}
\label{fig:penCleaning}
\end{figure}

\Cref{fig:copperCleaning,fig:penCleaning} shows the cleaning efficacy of the plasma on copper and marker pen. Preliminary experiments showed a pure air plasma was not ideal for cleaning and was always superseded in efficacy by an air-argon combination.

\FloatBarrier
\subsection{Surface removal rates}
\begin{figure}[ht]
\centering
\includegraphics[width=0.75\textwidth]{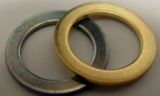}
\caption{A treated washer that has been exposed to a bright spot of air argon plasma for 8 minutes. An untreated gold washer is placed overlapping for contrast. Note the colourful oxides formed on the body of the washer and that the gold layer has been removed. }
\label{fig:etchedWasher}
\end{figure}

\begin{figure}[ht]
\begin{adjustwidth}{-\oddsidemargin-1in}{-\rightmargin}
\centering
\includegraphics[width=\paperwidth]{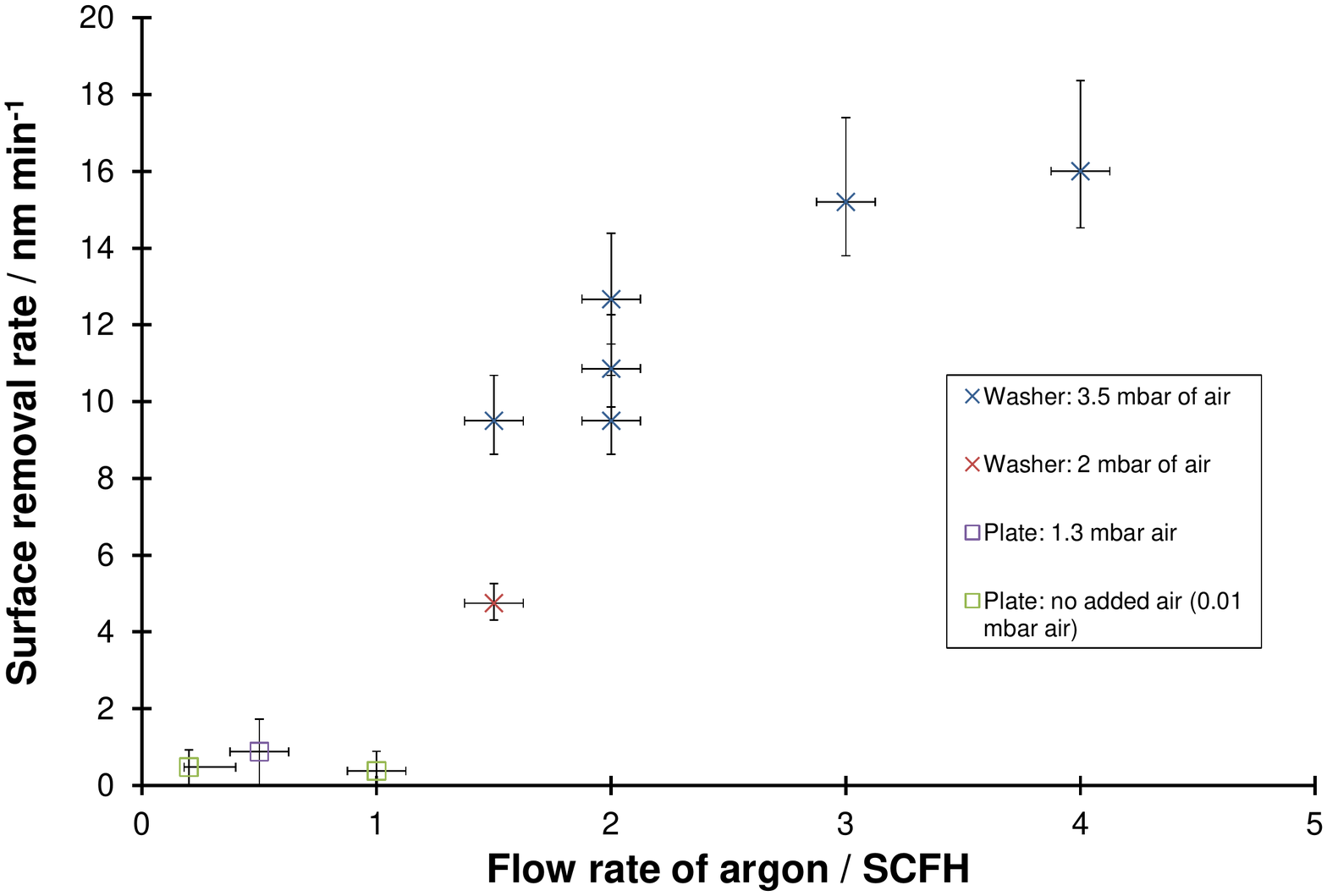}
\end{adjustwidth}
\caption{The estimated removal rates calculated from plasma exposure of the gold-plated washers (crosses) together with that calculated from the plates  (squares). It was possible to expose the gold-plated washers to higher flow rates as explained in \cref{subsec:methodEtch}. Errors were calculated for the plates by calculating the average and standard deviation of removal rate for each exposure type. Errors for gold-plated washers were calculated by accounting for the fact that the cleaning may have occurred between examinations and the uncertainty in gold layer thickness}
\label{fig:etchingRateGraphs}

\end{figure}

\cref{tab:goldPlates,fig:etchingRateGraphs} show the data obtained by XRF of the gold-coated aluminium plates and the calculated removal rates of gold from plates and washers. \Cref{fig:etchedWasher} shows such a washer after plasma cleaning.
\begin{table}[htbp]
\footnotesize
 \centering
    \begin{tabular}{|p{2cm}|p{2cm}|p{1.8cm}|p{1.5cm}|p{1.2cm}|p{1.2cm}|p{1.8cm}|}
    \toprule
    \textbf{Conditions} & \textbf{Gold removed \textbackslash nm} & \textbf{Exposure Time\textbackslash min} & \textbf{Pressure \textbackslash mbar} & \textbf{SCFH Argon} & \textbf{SCFH Air} & \textbf{Removal rate \textbackslash \unit[]{$\text{nm min}^{-1}$}} \\
    \midrule
    \multirow{3}{2cm}{Low pressure argon} & 16.7  & 18    & 0.17  & 0.2   & 0     & 0.93 \\
          & 10.0  & 36    & 0.18  & 0.2   & 0     & 0.28 \\
          & 3.3   & 54    & 0.18  & 0.2   & 0     & 0.06 \\
\bottomrule
    \multirow{3}{2cm}{Higher flow argon} & 16.7  & 18    & 4.9   & 1     & 0     & 0.93 \\
          & -3.3  & 36    & 5.15  & 1     & 0     & -0.09 \\
          & 16.7  & 54    & 5.2   & 1     & 0     & 0.31 \\
\bottomrule
    \multirow{3}{2cm}{Higher flow air-argon } & 26.7  & 18    & 4.5   & 0.5   & 0.5   & 1.48 \\
          & 10.0  & 36    & 4.5   & 0.5   & 0.5   & 0.28 \\
			& & & & & &\\
\bottomrule
    \bottomrule
    \end{tabular}%
  \caption{Tabulated results of the gold-coated aluminium plates experiment. The 8 plates were subjected to different conditions and the removal rate was calculated from the amount removed and time exposed. The XRF data showed that very little gold had been removed and longer exposures had, unexpectedly, not necessarily led to more removal. All exposures were initially pumped down to \unit[$(1\pm 0.5) \times 10^{-2}$]{mbar} of air. The amount of gold removed is calculated from the average of 3 XRF measurements at the centre of the plate before and after treatment. The error in pressure is 1 significant figure and in flow rate is \unit[$\pm 0.1$]{SCFH}.}
  \label{tab:goldPlates}%
\end{table}%

\FloatBarrier
\FloatBarrier
\clearpage \subsection{Finesse}
\begin{figure}[ht]

\centering
\subfloat[][670 nm]{\label{subfig:finesse670}\includegraphics[width=0.725\textwidth]{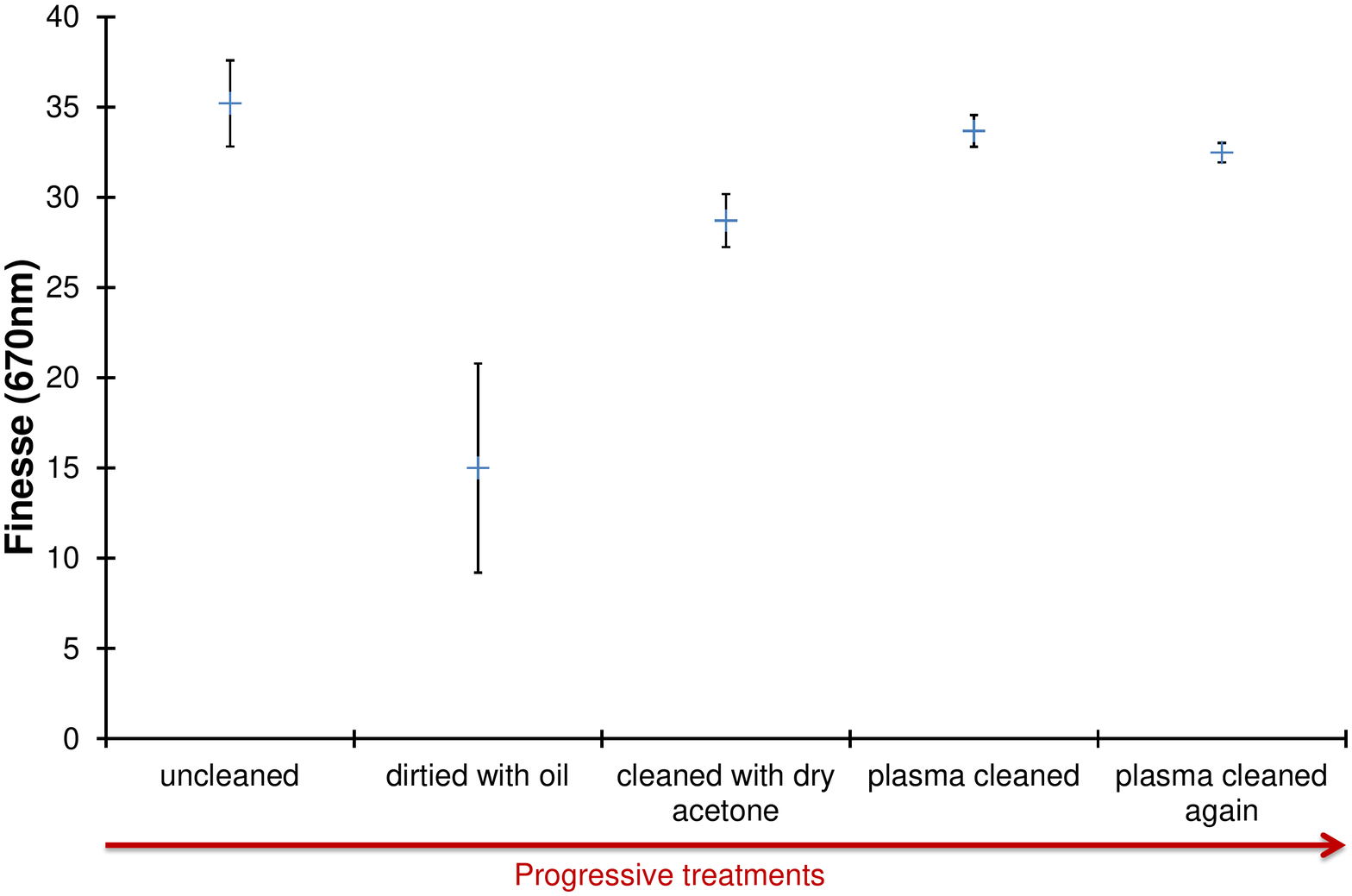}}
\hspace{0.01\textwidth}
\subfloat[][780nm]{\label{subfig:finesse780}\includegraphics[width=0.725\textwidth]{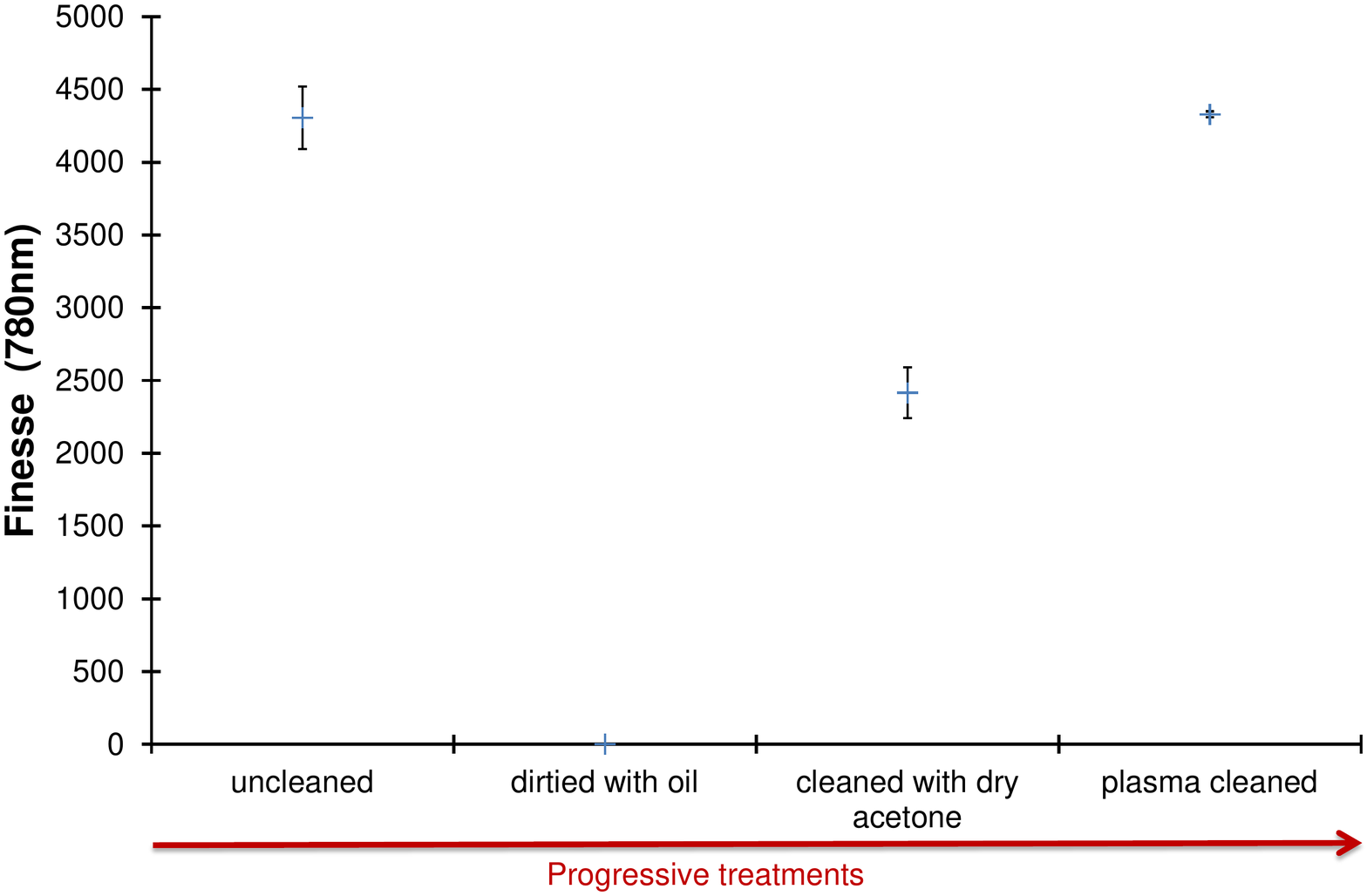}}
\label{fig:finesse}
\caption{Variation in the finesse of an optical cavity after different treatments. See \cref{subsec:methodFinesse} for more information on how these treatments were performed. The measurements were done at a wavelength of \unit[670]{nm} where the predicted $\mathscr{F}\approx 30 $ and at \unit[780]{nm} where $\mathscr{F}\approx 30,000$. Errors were calculated from the standard deviation of multiple measurements of peak width and separation.}
\end{figure}

\Cref{subfig:finesse670,subfig:finesse780} show how the measured finesse varied under different treatments at different wavelengths.

\section{Discussion}\label{sec:discussion}
\subsection{Qualitative observations of the plasma system}\label{subsec:discQualitativePlasma}
The observed plasma glow discharge, as seen in \cref{fig:plasmaColours,fig:pressuresPlasma}, confirms a basic level of functionality of the system, as predicted by the theoretical mechanism in \cref{sec:theoretical}. This Section discusses features of the observed plasma and qualitative observations of cleaning are discussed in \cref{subsec:discusQualitativeClean}. The removal rate and optical cavity experiments are discussed in \cref{subsec:discussEtch,subsec:discussFinesse} respectively. 

It is in agreement with the theoretical model that at sufficiently low pressures (\unit[$\leq 5 \times 10^{-3}$]{mbar}) the plasma did not ignite. At these pressures, electrons are diffusing to the walls faster than they are being created. It would be expected that the spectrum from the plasma is different depending on the gas used which explains the difference in colour upon the introduction of argon. The observation of the patch like structure formed in the air-argon plasma (\cref{fig:argonAirPlasma}) is caused by the non-uniformity of the MW field. This non-uniformity is well known and is why MW ovens contain a turntable to rotate food through hot spots. Thus, these patches are actually imaging the three dimensional standing wave pattern of the MW field --- a task that is regarded as difficult for MW oven engineers \cite{PhysOfMW}. It is unlikely that these patches are caused by the gas flow or glass bowl due to the symmetry of the setup compared to the asymmetry of the patches. In the dim regions between patches of bright plasma, it is likely that the field has interfered deconstructively and is insufficient to cause a breakdown (\cref{fig:critcalValSurf}, \cref{fig:argonAirPlasma}). It may seem disconcerting that the argon plasma is ignited uniformly, considering that the ionisation energy of argon is greater than that of the principle components of air. However, our analysis only considered a single pure gas in the vessel and the presence of a gas mixture (in our case air) causes substantial complications. It is likely in the air-argon system that the oxygen and nitrogen are the active plasma species with argon acting as a buffer gas, making it harder to ignite. For a thorough treatment of gas mixture plasmas, see \cite{MwBreakdownBook}. The plasma looks brighter near the magnetron source where the electric field is expected to be strongest and hence, the highest removal rates are expected in this region.

The reason the plasma appears to hug the inside of the glass bowl is explained by appealing to the plasma parameters defined in \cref{subsec:plasmaPararameters}. Using these definitions, a nominal value \cite{MWPlasmaAdvantages} of $n_e\sim \unit[10^{18}]{m^{-3}}$ gives $\omega_p = 2 \pi \times \unit[9]{GHz} \gg \omega $ and hence a decay length $\alpha^{-1}=\unit[5]{mm}$. Thus, the E-field decays by $e^{-1}$ over this length scale and rapidly the amplitude will be too small to cause ignition (\cref{fig:critcalValSurf}). This is in excellent agreement with the thin layer of plasma observed around the boundary of the chamber. Unfortunately, this would mean that the vast majority of our active cleaning plasma particles exist in a small region near the boundary and hence cleaning will occur primarily in this region. This was confirmed most easily by tests with copper sheets in which cleaning was rapid when the copper was near the glass but negligible or slow when central.

The labyrinthine pattern (\cref{fig:electricLines}) occurs at high flow rates, however, it is more likely that it is due to the correlated increase in pressure (\cref{fig:pressureFlow}). As the pressure increases we would expect the $N_\text{d}$, as defined in \cref{subsec:breakdownCriterion}, to increase, meaning our criterion for breakdown is more strongly satisfied. However, at sufficiently high pressure the \textit{optimum breakdown regime} is no longer valid \cite{MwBreakdownBook} and rapid liberation of electrons brings us into the arcing regime. The onset of this process has been thoroughly studied \cite{glowToArc,glowtoArc2,glowToArc3} and can be shown to occur at sufficiently high pressures. Hence, as the flow rate increases, causing the pressure to increase, the system transitions to the arcing regime and we see the pattern in \cref{fig:electricLines}. To predict the pressure this occurs at requires analysis of the Townsend avalanche and depends on the gas and ignition mechanism. This explains why it occurs at significantly different pressures for argon compared to air-argon plasmas. The arcing regime was shown to have negligible cleaning effect. It could possibly be prevented if the vacuum pumping speed was increased, such that at high flow rates the pressure was not changed significantly. This would involve increasing the diameter of the inlet piping and would require a new design of the baseplate (\cref{fig:basePlate}). High flow rates are desirable as they should lead to higher removal rates and a reduction in redeposition of sputtered material \cite{LVPROJ}. The onset of the arcing regime prevented us cleaning at the highest flow rates possible with the system.

\subsection{Qualitative observations of cleaning on copper and marker pen}\label{subsec:discusQualitativeClean}
From \cref{fig:penCleaning,fig:copperCleaning} it is clear the plasma is having a cleaning effect. A pure argon plasma cleaned the copper oxide layer to show a shinier, paler body of pure copper in \unit[1\text{--}2]{min} depending on location and flow rates. It was found that, in general, higher flow rates of argon cleaned more quickly, but the rapid onset of the arcing regime when pure argon was used prevented accurate quantifying of this result. The copper sheets were placed vertically and near the glass boundary for cleaning. When they were placed centrally negligible cleaning occurred, as expected due to the plasma only being active within close proximity to the glass boundary. Furthermore, when they were placed flat along the metal baseplate the cleaning rate was dramatically reduced even near the glass boundary. This is expected due to the fact that the parallel component of the electric field is continuous over a boundary and hence must go to zero at the baseplate. Thus, as the electric field goes to zero, we expect the density of active plasma species to decrease dramatically and hence a significantly smaller cleaning effect for objects placed flat along the baseplate. This could be utilised to gently clean sensitive electronic devices that cannot be exposed to large E-fields. If air is introduced into the plasma with a copper substrate a thick dark layer is produced on the copper due to reaction with the active oxygen particles in the air plasma.

When cleaning marker pen it was noted that the air-argon plasma was dramatically more effective than a pure argon plasma. This is explained by the reactive oxygen free radicals breaking down the complex hydrocarbon bonds found in ink chemicals. This highlights the fundamental difference between using air-argon and pure argon plasmas; for pure argon the cleaning is effectively caused only by sputtering as argon ions are unlikely to react with contaminants. However, for air-argon there is a significant quantity of more reactive free radicals (notably oxygen) and hence etching is possible and can be very effective in removing contaminants. Nevertheless, it is unsuitable to introduce air if the contaminants are metallic, or there could be unwanted reactions with oxygen. In this case, the chemically inert argon plasma is more suitable. It should be noted that the marker pen was only cleaned in the bright spots of the plasma (\cref{fig:penCleaning}) and this highlights that these regions have a high density of active plasma species and the dim regions do not clean as effectively.

\subsection{Quantification of surface removal rate}\label{subsec:discussEtch}

The data obtained from XRF of the gold-coated aluminium plates after plasma treatment (\cref{tab:goldPlates}) showed that for most of the plates very little gold had been removed. Furthermore, repetitions under the same conditions but with longer exposure did not necessarily lead to a larger removal. Thus, from this data it would be concluded that the plasma, under these conditions, had an insignificant effect on the removal of gold. This is quite contrary to the evidence obtained from cleaning of copper oxide, marker pen and gold-plated washers. For the plates only low flow rates were feasible in order to maintain a uniform distribution of plasma over the gold surface. At such low flow rates it is possible we were limited by low sputtering rates and redeposition of sputtered material. If one can consider the flow rate negligible, then a removed contaminant is expected to be redeposited on average a mean free path, $l$, away from its removal site. Under typical conditions $l\sim\unit[0.5]{mm}$ and hence at such low flow rates the gold may simply be getting redeposited on the surface. This could explain the negative removal rate measured for one sample. The XRF measurement was only done in one location (centre) of each plate and hence would not detect if the gold layer had been redistributed. Furthermore, it may be that the energy required for removal of a gold particle from its surface is more significant than that of a copper oxide which was removed under similar conditions. Discussion after the experiment with \textit{Modern Metallic Finishes Ltd} highlighted that a small quantity of nickel is added to their gold plating to make it harder to remove. The plates did appear visually different before and after treatment but this may be due to the removal of surface contaminants rather than gold.

The washers with gold plating of a known thickness were exposed to much higher flow rates and as shown in \cref{fig:etchingRateGraphs}, removal rates were found to be between \unit[5--15]{\nmmin} which is in reasonable agreement with the cleaning rates of copper oxide and marker pen. The graph shows a trend of faster removal rates with higher flows which is promising as at higher flow rates we expect lower redistribution of sputtered material. Furthermore it appears that if more air flows into the system the removal rate is faster. This may be due to an additional etching effect and the higher energy of oxygen free radicals. The downside of more air is that the bright spots become significantly smaller.

\subsection{Finesse of an optical cavity}\label{subsec:discussFinesse}
The purpose of measuring the finesse, $\mathscr{F}$, after different treatments was to quantify an improvement of an optical cavity upon plasma cleaning, compared to wet chemical techniques and thus to assess the applicability of plasma cleaning to optics. This experiment has direct relevance to cavity QED ion trapping experiments, which use optic fibres with the same lining as these mirrors to trap and excite ions simultaneously. For these experiments a high $\mathscr{F}$ is critical and any cleaning technique to improve this value would be useful. It was significantly easier to take measurements at the alignment wavelength of $\lambda=\unit[670]{nm}$ where the signal was stable to environmental perturbations and the predicted value of $\mathscr{F}$ was 30. At the resonant wavelength of $\lambda=\unit[780]{nm}$, $\mathscr{F}$ is predicted to be 30,000 and the cavity system is highly unstable to environmental perturbations such as nearby motion or air movement. This was seen by rapid movement of the signal on the oscilloscope which made it harder to find the correct alignment and make accurate measurements of the peak width. 

At $\lambda=\unit[670]{nm}$ (\cref{subfig:finesse670}) the $\mathscr{F}$ was found to be slightly higher than predicted and fell by greater than a factor of 2 after one mirror was dirtied. The Lorentz form of the peak looked significantly degraded after dirtying. Cleaning of the mirror with dry acetone had a significant effect on improving $\mathscr{F}$ but the cavity did not reach value of $\mathscr{F}$ attained before dirtying. This value of $\mathscr{F}$ was reached, within error, after plasma cleaning. This strongly suggests that the plasma cleaning has removed more contaminants than wet chemical cleaning and that this has had a positive effect on $\mathscr{F}$. The mirror was then plasma cleaned again to see if additional cleaning would affect $\mathscr{F}$. Addition plasma exposure very slightly reduced $\mathscr{F}$ which might suggest that excessive plasma treatment can damage surfaces perhaps by causing surface defects or ion implantation \cite{LVREF7}. Considering the mirror had been exposed for \unit[250]{s} and with an approximate removal rate of \unit[10]{\nmmin} it was unlikely further plasma cleaning would have a beneficial effect.

At $\lambda=\unit[780]{nm}$ (\cref{subfig:finesse780}) a very similar trend is seen and the difference in $\mathscr{F}$ between acetone and plasma treatment is even more significant ($\mathscr{F}\sim 2500$ to $\mathscr{F}\sim 4250$). It was noted that the order of magnitude of the measured $\mathscr{F}$ was lower than the predicted value. This could be due to a poor alignment of the cavity system, or perhaps a problem with the mirror surfaces. Also, as one mirror was attached to the piezo and not able to be plasma cleaned, it was left untreated throughout the experiment and perhaps contamination of that mirror caused the lower value of $\mathscr{F}$. At this wavelength dirtying the mirror caused the signal to be destroyed completely and hence $\mathscr{F}=0$ was given for this state. The results are very promising in that they show a return to pre-contaminated values after a \textit{catastrophic contamination event} and an increase in $\mathscr{F}$ compared to wet chemical cleaning alone. The conditions used to plasma treat the mirror (as described in \cref{subsec:methodFinesse}) were chosen to provide a reasonable removal rate over the entire region of the mirror and were based on the results discussed in \cref{subsec:discusQualitativeClean,subsec:discussEtch}.

\section{Conclusion}\label{sec:conclusion}

In this experiment a plasma cleaning system was constructed by adapting a conventional MW oven. Copper oxide and marker pen were cleaned in the system which demonstrated its functionality. By examining the rate of removal of the gold on gold-plated washers, an estimate of the removal rate was found to be \unit[5--15]{\nmmin} which can be compared to the nominal value of a commercial RF plasma cleaner (\pounds 10,000) of \unit[30]{\nmmin} \cite{LVREF7}. An experiment was performed by measuring the finesse of an optical cavity after different treatments. Plasma treatment after cleaning with dry acetone gave a finesse approximately twice that attainable with dry acetone alone. This suggests our system is suitable for cleaning atomic scale contaminants to a higher degree than is possible with wet chemical techniques. Furthermore, it suggests it may be useful in cleaning optical devices used in ion trapping experiments where minimising levels of surface contaminants is critical.

A primary aim of this experiment was to assess if this system and technique would be valid for treating ion trap electrodes, for which anomalous heating rates have been reported and are thought to be due to small quantities of contaminants on the electrode surfaces as explained in \cref{sec:intro}. Recently Wineland et al. \cite{R15} have shown a reduction of two orders of magnitude in heating rates after trap electrodes were cleaned using an argon ion beam. This technique would likely have very similar results to our plasma cleaning system where an argon plasma can easily be produced. Furthermore, in the same paper, they confirm using Auger Electron Spectroscopy that the main surface contamination is by carbon and hence a suitably chosen gas combination could be even more effective. These results are very promising for using this technique on trapped ion quantum computers and it is hoped that our laboratory will attempt to use the system on upcoming ion trap experiments. However, implementing our MW based system in-situ is not yet possible and would require a revision of design.

Two main factors limited our cleaning potential and could be focal points for further work on this system. Firstly the plasma does not emanate far beyond the glass boundary due to the skin depth of the plasma as discussed in \cref{subsec:discQualitativePlasma}. Whilst, for most of the small objects that were cleaned this did not cause a problem, for large surfaces it becomes difficult to ensure a uniform cleaning rate over the surface. This effect could be seen as an advantage as it means by placing objects more centrally they are exposed to fewer active plasma species and hence experience a more gentle cleaning effect. Work done on MW plasma shows it is possible to circumvent this problem by using strong magnetic fields which alter the plasma and E-field distributions to ensure a more homogeneous plasma \cite{R21}. Secondly, it was seen that at high flow rates a transition to the arcing regime occurred which effectively stopped cleaning. In the glow discharge regime it was seen that high flow rates led to a higher removal rate and greater cleaning effect. Thus, it was detrimental that even higher flow rates could not be attained in the glow discharge regime. If the pumping speed could be increased, then it is feasible that we could increase the flow rates whilst maintaining a lower pressure. From the vacuum set up of the system the only significant way to increase the pump speed would be to increase the diameter of the vacuum outlet from the baseplate. This would require a redesign of the baseplate and most likely involve the vacuum outlet coming from underneath rather than the side, as it is currently. Nevertheless, the home-built plasma cleaning system has shown itself to be effective under suitable conditions and further use will highlight its applicability for optics and Paul trap electrodes.

\section*{Acknowledgements}

I would like to thank Professor Michael K\"ohl, Matthias Steiner and Hendrik Meyer (\textit{University of Cambridge}) for their regular help, advice and encouragement throughout the project. Thanks is also extended to the PhD students and postdoctoral research staff of the \textit{Atomic, Mesoscopic and Optical Physics} group, whose discussion and advice was appreciated.
Wayne Love (\textit{Microwave Research and Applications Inc.}) and Professor Michael Vollmer (\textit{University of Applied Sciences, Brandenburg}) are thanked for their technical advice on microwave generation and absorption. Professor Roy Tumlos (\textit{University of the Philippines Diliman}) is thanked for his advice on modern techniques in microwave plasmas.

\bibliography{projectBib}
\bibliographystyle{ieeetr}

\appendix
\newcommand{\appsection}[1]{\let\oldthesection\thesection
  \renewcommand{\thesection}{Appendix \oldthesection}
  \section{#1}\let\thesection\oldthesection}

\appendixpage
\appsection{MW oven customisations}\label{app:MWCustomisation}

The choice of the MW oven involved the consideration of several factors. The dielectric material inside the cavity effects the field, power and phase of the generated MW radiation and hence the MW field inside the cavity. This is why the MW field is often described as dynamic with the load. To initiate a plasma glow discharge, high fields or locations of high constructive interference (HCI), are required within the argon gas (\cref{sec:theoretical}). MW ovens usually use a rotating turntable to rotate food through areas of HCI to prevent one region being overheated. If the argon vessel is smaller than and contained within a region of HCI, then an ideal setup could be attained where the vessel is evenly exposed to MW radiation. However, if the regions of HCI are smaller than the vessel and exist outside of it, the conditions would be unsuitable. Furthermore, a rotating turntable introduces a motor mechanism that could obstruct the modifications or baseplate. A subset of MW ovens use a rotating antenna at the entrance to the cavity to mix MW radiation modes and cause the areas of HCI to move dynamically. This would be desirable unless the mode stirrer simply moved areas of HCI out of our vessel altogether. Most modern MW ovens that use the rotating antenna mechanism also have a rotating turntable, presumably as consumers feel more at ease if they can see their food rotating! Finally, a MW oven with an inverter type power supply should not be used as these chop the anode current supplying the magnetron into high frequency chunks and gives less suitable HCI for plasma initiation. A \textit{Samsung MW 76N-B/XEU} \unit[800]{W} MW oven was chosen as it was economic and was simply wired, leaving the left hand side free from electronics and easy to modify. It did not have a rotating antenna.

\begin{figure}[ht]
\centering
\subfloat[][Interior with motor removed and drilled holes]{\label{subfig:removedMotor}\includegraphics[width=0.4\textwidth]{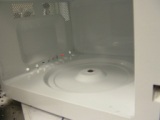}}
\hspace{0.01\textwidth}
\subfloat[][Exterior view of holes and baseplate inserted]{\label{subfig:holes}\includegraphics[width=0.4\textwidth]{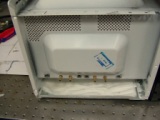}}
\hspace{0.01\textwidth}
\subfloat[][Baseplate grounded using screws]{\label{subfig:holesAndScrews}\includegraphics[width=0.4\textwidth]{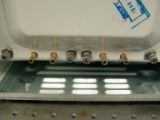}}
\hspace{0.01\textwidth}
\subfloat[][Hoses added ]{\label{subfig:holesAndPipes}\includegraphics[width=0.4\textwidth]{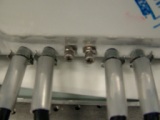}}
\hspace{0.01\textwidth}
\subfloat[][Interior wire of MW oven showing heat monitoring setup]{\label{subfig:interiorWire}\includegraphics[width=0.4\textwidth]{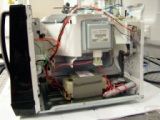}}
\hspace{0.01\textwidth}
\subfloat[][Washers on hose connectors help ground from the inside]{\label{subfig:hoseWashers}\includegraphics[width=0.4\textwidth]{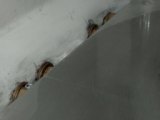}}
\hspace{0.01\textwidth}
\subfloat[][The complete system]{\label{subfig:Complete}\includegraphics[width=0.4\textwidth]{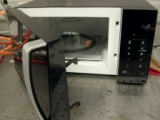}}
\caption{The modifications performed to the MW oven to make it suitable for plasma generation. These figures highlight how the MW oven was adapted and are referred to in the text of \ref{app:MWCustomisation}}
\end{figure}

The turntable was removed from the MW oven and the protruding part of the motor mechanism was detached to prevent it abrading the underside of the baseplate (\cref{subfig:removedMotor}). The baseplate, as described in \cref{sec:method,fig:basePlate}, lay inside the cavity in order to allow gas to flow into and be removed from the vessel. It was water-cooled to \unit[7]{\degree C} using a simple U-shaped path through its interior. M6 threads were added to allow mounting of substrates if required. A further four additional threads were added to the side where the hose connectors emanated to allow the base plate to be secured and grounded to the MW oven. Eight holes were then drilled into the side of the inside chamber to allow the baseplate to sit inside with the hose connectors passing through (\cref{subfig:holes}). The paint was removed from the regions surrounding these holes to allow for better grounding. Small copper sheets with central holes were used as washers to ensure a good connection of the screws and hose connectors to the MW oven (\cref{subfig:holesAndScrews,subfig:hoseWashers}). Piping to the hose connectors was secured and connected the baseplate to the necessary external sources (\cref{subfig:holesAndPipes}). The secondary external metal casing was then cut to allow these pipes to exit from the system.

Once the baseplate was secured, the system was tested to ensure functionality by heating a beaker of water. Arcing was occurring between the baseplate and regions where paint had accidently been scratched off. Teflon or Kapton Amide sheets were used to cover these regions and effectively prevented arcing. The system was checked for MW radiation leakage using an RF spectrum analyser and comparing the signals at various fixed locations to the pre-modification levels. Foil and aluminium sheets were used to cover the modified regions and prevent leakage into the electric circuitry or out of the system.

Inverted glass kitchen bowls were found to be a strong, cheap and effective container for the plasma system, with various sizes readily available. The rims of the bowls were machined with a diamond lathe to ensure they were flat. To connect the bowls to the baseplate a \unit[2]{mm} sheet of Viton Rubber  was cut into a ring and placed between the baseplate and bowl edge which provided a tight seal; attempts using vacuum were less effective. Viton can withstand approximately \unit[200]{\degree C} and burning occurred only in a pure air plasma when water cooling was not used.

Gas flow into the chamber via the baseplate was controlled using two flowmeters (\textit{Dwyer Instruments RMA-5 with top mounted valve for vacuum application}) connected via a T-junction to allow the admission of argon (99.99\% pure) and air into the system. Oxygen would have been more desirable than air but was not used for safety reasons. The flowmeters were found to be the dominant source of leaks into the system and copious application of vacuum grease rectified this issue.

Running a MW oven empty is not recommended by manufacturers as reflected power couples back into the magnetron source and causes overheating which significantly reduces the lifetime of the device. Thus, a PT-100 temperature dependent resistor was fixed to the outside of the magnetron (\cref{subfig:interiorWire}). This allowed us to monitor this temperature which was kept at a reasonable level by running the MW oven on a lower power (which simply switches off the magnetron for a fraction of the cycle whilst the fan still operates) and using the ventilation function of the device to simply run the fan for a specified time. MW absorbing materials could also be placed within the chamber to reduce overheating, however, this was found to have a negligible effect unless beakers of water were used.

\end{document}